\documentclass[11pt]{article}

\usepackage{latex8}
\usepackage{times}

\usepackage{amssymb}
\usepackage{amsmath}
\usepackage{graphics}
\usepackage{color} 
\usepackage{fullpage}

\ifx\pdftexversion\undefined
  \usepackage[dvips]{graphicx}
\else
  \usepackage[pdftex]{graphicx}
\fi

\newtheorem{theorem}{Theorem}
 \newtheorem{lemma}[theorem]{Lemma}
\newenvironment{proof}[1][Proof]{\noindent\textbf{#1.} }{\hfill $\Box$\\[2mm]} %\rule{0.5em}{0.5em}\\}
\def\lf{\tiny}

\def\nnll{\refstepcounter{linenumber}\lf\thelinenumber}
\newcounter{linenumber}

\def\Time{\mathbb{T}}

\def\D{\ensuremath{\mathcal{D}}}

\def\A{\ensuremath{\mathcal{A}}}

\def\R{\ensuremath{\mathcal{R}}}

\def\M{\ensuremath{\mathcal{M}}}

\def\E{\ensuremath{\mathcal{E}}}

\def\Nat{\ensuremath{\mathbb{N}}}

\def\fd{failure detector}

\newcommand{\correct}{\mathit{correct}}
\newcommand{\faulty}{\mathit{faulty}}
\newcommand{\true}{\mathit{true}}

\newcommand{\remove}[1]{}

\newcommand{\id}[1]{\mbox{\textit{#1}}}% for identifiers in code
\newcommand{\ignore}[1]{}

\begin{document}

\bibliographystyle{plain}

\title{Simple CHT: \\ 
A New Derivation of the Weakest Failure Detector for Consensus}

\author{
Eli Gafni\\
\\
\normalsize UCLA
%, Los Angeles. 
%3731F Boelter Hall, UCLA, LA. CA. 90095, USA. eli@ucla.edu
%}
\and
Petr Kuznetsov\\ 
\\
\normalsize T\'el\'ecom ParisTech
%Campus, E1.4, D-66123 Saarbr\"ucken, Germany. Fax: +49 681 9325 697, e-mail: pkouznet@mpi-sws.mpg.de
%}
%\\
%\\
}

%%%%%%%%%%%%%%%%%%%%%%%%%%%%%%%%%%%%%%%%%%%%%%%%%%%%%%%%%%%%%%%%%%%%%%%%%%%%%%%%
\date{}
\maketitle

\begin{abstract}
The paper proposes an alternative proof that $\Omega$, an oracle that outputs 
a process identifier and guarantees that eventually the same 
correct process identifier is output at all correct processes,
provides minimal information about failures for solving consensus 
in read-write shared-memory systems: 
every oracle that gives enough failure information to solve consensus 
can be used to implement $\Omega$.

Unlike the original proof by Chandra, Hadzilacos and Toueg (CHT), the proof
presented in this paper builds upon the very fact that $2$-process wait-free 
consensus is impossible. 
Also, since the oracle that is used to implement
$\Omega$ can solve consensus, the implementation is allowed 
to directly access consensus objects. 
As a result, the proposed proof is shorter and conceptually simpler than the original one.
\end{abstract}

%\begin{center}
%\textbf{Keywords:} $k$-impossible tasks, synchrony assumptions, failure detectors, $k$-set agreement
%\end{center}

\section{Introduction}

The presence of faults and the lack of synchrony
make distributed computing interesting. 
On the one hand, in an \emph{asynchronous} system which assumes no bounds on
communication delays and relative processing speeds, even a basic form of 
non-trivial synchronization (\emph{consensus}) 
%using conventional read-write shared memory
is impossible if just one process may fail by crashing~\cite{FLP85,LA87}.
On the other hand, in a \emph{synchronous} system, 
where the bounds exist and known a priori and every failure can be
reliably detected, 
every meaningful fault-tolerant synchronization problem becomes solvable. 
%Thus, in synchronous systems, there is no distinct difference between 
%the classical theory of computability and its distributed counterpart.
The gap suggests that the amount of information about failures %needed to solve a distributed computing problem 
is a crucial factor in reasoning about fault-tolerant solvability of distributed computing problems.
%into equivalence classes 
%based on the synchrony assumptions required to solve them.

Chandra and Toueg proposed \emph{failure detectors} as
a convenient abstraction to describe the failure information.
Informally, a failure detector is a distributed oracle that 
	provides processes with hints about failures~\cite{CT96}. 
The notion of a \emph{weakest failure detector}~\cite{CHT96} 
	captures the exact amount of synchrony needed to solve a given problem:
	$\D$ is the weakest failure detector for solving $\M$ if 
(1) $\D$ is sufficient to solve $\M$, i.e., there exists an algorithm that solves $\M$ using $\D$,  and (2) any failure detector $\D'$ that is sufficient to solve $\M$
	provides at least as much information about failures as $\D$ does,
	i.e., there exists a \emph{reduction} algorithm that 
	extract the output of $\D$ using the failure information provided by $\D'$. 
   
This paper considers a distributed system % read-write shared-memory model 
	in which $n$ crash-prone processes communicate using atomic reads and writes 
	in shared memory.
In the (binary) consensus problem~\cite{FLP85},   
	every process starts with a binary input and
	every correct (never-failing) process
	is supposed to output one of 
	the inputs such that no two processes output
	different values. 
Asynchronous (wait-free) consensus is known to 
	be impossible~\cite{FLP85,LA87}, as long as at least one process
	may fail by crashing.   
Chandra et al.~\cite{CHT96} showed that the ``eventual leader'' failure detector 
	$\Omega$ is necessary and sufficient to solve consensus.
The failure detector $\Omega$ outputs, when queried, a process identifier, 
	such that, eventually,  the same correct process identifier is output 
	at all correct processes. 

The reduction technique presented in~\cite{CHT96} is very interesting in its own right,
	since it not only allows us to determine the weakest failure detector for consensus, 
	but also establishes a framework for determining the weakest failure detector for 
	any problem.
Informally, the reduction algorithm of~\cite{CHT96} works as follows.
Let $\D$ be any failure detector that can be used to solve consensus. 
Processes periodically query their modules of $\D$, exchange
	the values returned by $\D$, and arrange the accumulated 
	output of the failure detector in the form 
	of ever-growing directed acyclic graphs (DAGs). 
Every process periodically uses its DAG as a stimulus for simulating multiple 
	runs of the given consensus algorithm.  
It is shown in~\cite{CHT96} that,
	eventually, the collection of simulated runs will include
	a \emph{critical} run in which a single process $p$ ``hides'' the 
	decided value, and, thus, no extension of the run can reach a decision 
	without cooperation of $p$.
As long as a process performing the simulation observes a run that
	the process suspects to remain critical, 
	it outputs the ``hiding'' process identifier of the ``first'' such run
	as the extracted output of $\Omega$.  
The existence of a critical run and the fact that the correct 
	processes agree on ever-growing prefixes of simulated runs imply 
	that, eventually, the correct processes will always 
	output the identifier of the same correct process. 

Crucially, the existence of a critical run is established in~\cite{CHT96} using 
	the notion of \emph{valence}~\cite{FLP85}:
	a simulated finite run is called $v$-valent ($v\in\{0,1\}$) 
	if all simulated extensions of it decide $v$. 
If both decisions $0$ and $1$ are ``reachable'' from the finite run, then 
	the run is called bivalent.
Recall that in~\cite{FLP85}, the notion of valence is used to derive a critical run,
	and then it is shown that such a run cannot exist in an asynchronous system,
	implying the impossibility of consensus.
In~\cite{CHT96}, a similar argument is used to extract the output of $\Omega$ in 
	a partially synchronous system that allows for solving consensus. 
Thus, in a sense, the technique of \cite{CHT96} rehashes 
	arguments of~\cite{FLP85}.
It is challenging to find a proof that $\Omega$ is necessary to solve consensus 
	building upon the very fact that $2$-process wait-free consensus is impossible.

This paper addresses this challenge. 
It is shown that $\Omega$ is necessary to solve consensus using
	the very impossibility of $2$-process wait-free consensus, 
	without ``opening the box'' and 
	considering the problem semantics. 
The resulting proof is shorter and simpler that the original 
	proof of~\cite{CHT96}. 

On the technical side, the paper uses two fundamental results and one observation.
First, the technique of Zieli\'nski~\cite{Zie08}  
	that allows us to construct, given an algorithm $\A$ that uses a failure detector $\D$, 
	an \emph{asynchronous} algorithm $\A'$  that simulates 
	runs of $\A$ using a static sample of $\D$'s output (captured in a DAG), 
	instead of the ``real'' output of the failure detector.
The live processor set that run $\A'$ may be different than the live set of processor implied
by the sample of $\D$.
Therefore, the asynchronous algorithm $\A'$ guarantees that every infinite simulated run 
	is \emph{safe} (every prefix of it is a finite run of $\A$),
	but not necessarily \emph{live} (some correct process may not 
	be able to make progress). 

Second, the paper also makes use of the BG-simulation technique~\cite{BG93b,BGLR01} that allows $k+1$ 
	processes simulate, in a wait-free manner, a \emph{$k$-resilient} 
	(with at most $k$ faulty processes) 
	run of $\A'$.
Using a series of consensus instances (provided by the algorithm $\A$ using $\D$),
	processes locally simulate the very same sequence of $1$-resilient 
	runs and eventually identify a ``never-deciding'' $1$-resilient run 
	of $\A'$.
Since $\A'$ is an asynchronous simulation of $\A$, a $1$-resilient run of $\A'$ 
	that includes infinitely many steps of each correct
	process should be deciding.
Thus, exactly one correct process appears 
	in the $1$-resilient never-deciding run only finitely often.    
To emulate $\Omega$, it is thus sufficient to output the process 
	that appears the least in that $1$-resilient run.
Eventually, all correct process agree on the same never-deciding $1$-resilient run, and 
	will always output the same correct process. 
The observation here is that a reduction algorithm  may directly access
	consensus objects, since it is given a failure detector 
	which can be used to solve consensus. 

%This technique can be seen as a generalization of earlier results in
%	which a necessary failure detector for solving a wait-free impossible 
%	problem was derived from the very fact that the problem is 
%	wait-free impossible~\cite{GHKLN07,Zie08}.	

%As a side effect, this paper gives an alternative 
%	proof that $\Omega$ is the weakest failure detector for solving consensus
%	in the read-write shared memory model.

%\paragraph{Roadmap.}
The rest of the paper is organized as follows.
Section~\ref{sec:model} describes the system model.
Sections~\ref{sec:nec} presents the reduction algorithm.
Section~\ref{sec:related} overviews the related work and Section~\ref{sec:discussion} 
	concludes the paper by discussing implications of the presented results.

%%%%%%%%%%%%%%%%%%%%%%%%%%%%%%%%%%

%%%%%%%%%%%%%%%%%%%%%%%%%%%%%%%%%%%%
\section{Model}
\label{sec:model}
%%%%%%%%%%%%%%%%%%%%%%%%%%%%%%%%%%%%

The model of processes communicating through read-write shared objects 
and using {\fd s} is based on~\cite{CHT96,GHKLN07,GK08,JT08}.
The details necessary for showing the results of this paper are described below. 

\subsection{Processes and objects}
A distributed system is composed of a set $\Pi$
of $n$ processes $\{p_1,\ldots,p_{n}\}$ ($n\geq 2$).
Processes are subject to \emph{crash} failures.   
A process that never fails is said to be \emph{correct}.
Processes that are not correct are called~\emph{faulty}.
Process communicate through applying atomic operations on a collection of 
	\emph{shared objects}.
In this paper, the shared objects are registers, i.e., they export 
	only conventional atomic read-write operations.
%The necessary parts of our results do not restrict the types of shared objects,
%	but, for simplicity of presentation we assume that the objects are \emph{deterministic},
%	i.e., the value returned by an object is solely determined 
%	by the object's state and the applied operation.

\subsection{Failure patterns and failure detectors}

A \textit{failure pattern} $F$ is a function from the time range $\Time=\{0\}\cup\Nat$ to 
	$2^{\Pi}$, where $F(t)$ denotes the set of processes that have crashed by time $t$. 
Once a process crashes, it does not recover, i.e., $\forall t: F(t) \subseteq F(t+1)$.
The set of faulty processes in $F$, $\cup_{t\in \Time} F(t)$, is denoted by $\faulty(F)$.
Respectively, $\correct(F)=\Pi-\textit{faulty}(F)$.
A process $p \in F(t)$ is said to be \textit{crashed} at time $t$.
An \emph{environment} is a set of failure patterns. 
This paper considers environments that consists of failure patterns 
in which at least one process is correct.

A \emph{failure detector history $H$ with range $\R$} is a function from
	$\Pi\times \Time$ to $\R$. 
$H(p_i,t)$ is interpreted as the value output by the \fd~module of process $p_i$ at time $t$.
A \textit{failure detector} $\D$ with range $\R_{\D}$ is a function that maps each failure 
pattern to a (non-empty) set of failure detector histories with range $\R_{\D}$. 
$\D(F)$ denotes the set of possible \fd~histories permitted by 
	\D~for failure pattern $F$.
Possible ranges of failure detectors are not a priori restricted.

\subsection{Algorithms}
An \textit{algorithm $\A$ using a failure detector $\D$} 
	is a collection of deterministic automata, one for each process in the system.
%	and an \emph{initial memory state}, i.e., 
%	the initial states of all shared objects used by the algorithm.
$\A_i$ denotes the automaton on which process $p_i$ runs the algorithm $\A$.
Computation proceeds in atomic \textit{steps} of $\A$. 
In each step of $\A$, process $p_i$ 

\begin{enumerate}

\item[(i)] invokes an atomic operation (read or write) on a shared object and receives a 
	response \emph{or} queries its failure detector module $\D_i$ 
	and receives a value from $\D$, and 

\item[(ii)] applies its current state, the response received from the shared object
	or the value output by $\D$ to the automaton $\A_i$ to obtain a new state.

%\item[(iii)] accepts an application \emph{input} %in $\I$ 
%	or produces (according to the automaton $\A_i$) an \emph{output}.
\end{enumerate}
%
%Since the algorithms and shared objects we consider  are deterministic, 
A step of $\A$ is thus identified by a tuple $(p_i,d)$, where $d$ is 
	the failure detector value output at $p_i$ during that step if $\D$ was queried,
	and $\bot$ otherwise.

If the state transitions of the automata $\A_i$ 
	do not depend on the failure detector values, the algorithm $\A$ 
	is called \emph{asynchronous}.
Thus, for an asynchronous algorithm, a step is uniquely identified by the process id.

%We assume that atomic objects are read-write registers: a \text{read}$()$ operation 
%	returns the value of the register, and a \text{write}$(v)$ operation 
%	changes the value to $v$. 

\subsection{Runs}
A \textit{state} of $\A$ defines the state of each 
	process and each object in the system. 
An \emph{initial state} $I$ of $\A$ specifies 
	an initial state for every automaton $\A_i$
	and every shared object. 

A \textit{run of algorithm $\A$ using a failure detector ${\D}$} in an environment $\E$  
is a tuple $R=\langle F,H,I,S,T \rangle$ where $F\in\E$ is a failure pattern, 
	$H\in {\D}(F)$ is a failure detector history, $I$ is an initial state of $\A$, 
	$S$ is an \textit{infinite} sequence of steps of $\A$
	respecting the automata $\A$ and the sequential specification of shared objects, 
	and $T$ is an \textit{infinite} list of
	increasing time values indicating when each step of $S$ has occurred,
	such that for  all $k\in\Nat$, if $S[k]=(p_i,d)$ with $d\neq \bot$, then 
	$p_i\notin F(T[k])$ and $d=H(p_i,T[k])$.

A run $\langle F,H,I,S,T \rangle$ is \emph{fair} 
	if every process in $\correct(F)$ takes infinitely many steps in $S$,	
	and \emph{$k$-resilient} if at least $n-k$ processes appear in $S$
	infinitely often.
A \emph{partial run} of an algorithm $\A$ is a finite prefix of a run of $\A$.

For two steps $s$ and $s'$ of processes $p_i$ and $p_j$, respectively,
	in a (partial) run $R$ of an algorithm $\A$, 
	we say that $s$ \emph{causally precedes} $s'$ if in $R$, and we write $s\to s'$, 
	if (1) $p_i=p_j$, and $s$ occurs before $s'$ in $R$, or 
	(2) $s$ is a write step, $s'$ is a read step, and $s$ occurs before $s'$ in $R$,
	or (3) there exists $s''$ in $R$, such that $s\to s''$ and $s''\to s'$.

\subsection{Consensus}
In the binary consensus problem, every process starts the computation with an 
	input value in $\{0,1\}$
	(we say the process \emph{proposes} the value), and 
	eventually reaches a distinct state associated with an output value in $\{0,1\}$
	(we say the process \emph{decides} the value).
An algorithm $\A$ solves consensus in an environment $\E$ if in every \emph{fair} run of $\A$ in $\E$, 
	(i) every correct process eventually decides,
	(ii) every decided value was previously proposed, and 
	(iii) no two processes decide different values.

Given a an algorithm that solves consensus, it is straightforward to implement an abstraction 
	\textsf{cons} that can be accessed with an operation \textit{propose}$(v)$ ($v\in\{0,1\}$)
	returning a value in $\{0,1\}$, and
	guarantees that every \textit{propose} operation invoked by a correct process eventually
	returns, every returned value was previously proposed, and no two different 
	values are ever returned.

\subsection{Weakest failure detector}
We say that an algorithm $\A$ using ${\D}'$ \emph{extracts} the output of $\D$ in $\E$,
	if $\A$ implements a distributed variable $\D\textit{-output}$ such that
	for every run $R=\langle F,H',S,T \rangle$ of $\A$ in which $F\in\E$,
	there exists $H\in\D(F)$ such that for all $p_i\in\Pi$ and $t\in\Time$,
	$\D\textit{-output}_i(t)=H(p_i,t)$ (i.e., the value of $\D\textit{-output}$ 
	at $p_i$ at time $t$ is $H(p_i,t)$).
We say that $\A$ is a \emph{reduction} algorithm. 
(A more precise definition of a reduction algorithm is given in~\cite{JT08}.)

If, for {\fd}s ${\D}$ and ${\D}'$ and an environment $\E$, 
	there is a reduction algorithm using $\D'$ that extracts the output $\D$ in $\E$, 
	then we say that \emph{${\D}$ is weaker than ${\D'}$} in $\E$.
%If $\D$ and $\D'$ are weaker than each other in $\E$, we say they  are equivalent in $\E$.

$\D$ is the weakest failure detector to solve a problem $\M$ (e.g., consensus) in $\E$ if there is an algorithm that 
	solves $\M$ using $\D$ in $\E$  and $\D$ is weaker than 
	any failure detector that can be used to solve $\M$ in $\E$.

%=====================================================================
\section{Extracting $\Omega$}
\label{sec:nec}
%=====================================================================

Let $\A$ be an algorithm that solves consensus using a failure detector $\D$.
The goal is to construct an algorithm that 
	emulates $\Omega$ using $\A$ and $\D$.
Recall that to emulate $\Omega$ means to output, at each time and at each process, 
	a process identifiers such that, eventually, the same correct process 
	is always output.

\subsection{Overview}
\label{sec:overview}

As in~\cite{CHT96}, the reduction algorithm of this paper uses  
	 failure detector $\D$ to construct 
	an ever-growing \emph{directed acyclic graph} (DAG) that
	contains a sample of the values output by $\D$ in the current run
	and captures some temporal relations between them.
Following~\cite{Zie08}, this DAG can be used by an 
	\emph{asynchronous} algorithm $\A'$ to simulate 
	a (possibly finite and unfair) run of $\A$.
Recall that, using BG-simulation~\cite{BG93b,BGLR01}, $2$ processes can
	simulate a $1$-resilient run of $\A'$. 
The fact that $1$-resilient $2$-process consensus is impossible 
	implies that the simulation
	must produce at least one "non-deciding" $1$-resilient run of $\A'$.

Now every correct process locally simulates all executions of BG-simulation 
	on two processes $q_1$ and $q_2$ that simulate
	a $1$-resilient run of $\A'$ of the whole system $\Pi$.  
Eventually, every correct process locates a never-deciding run of $\A'$ 
	and uses the run to extract the output of $\Omega$:
	it is sufficient to output the process that 
	takes the least number of steps in the  ``smallest'' 
	non-deciding simulated run of $\A'$.
Indeed, exactly one correct process takes finitely many steps in the 
	non-deciding $1$-resilient run of $\A'$:
	otherwise, the run would simulate a fair and thus 
	deciding run of $\A$.
%  
%Indeed, a never-deciding $k$-resilient run $R$ cannot be fair 
%	(otherwise, since $\A$ solves the task using $\D$, it would be deciding) and,
%	thus, it is sufficient to output any 
%	$n-k$ processes that appear infinitely often in $R$.

The reduction algorithm extracting $\Omega$ from $\A$ and $\D$ consists of two components that are running
        in parallel: the \emph{communication component} and the \emph{computation component}.
In the communication component, every process $p_i$ maintains the 
	ever-growing directed acyclic graph (DAG) $G_i$ by periodically querying its {\fd} module
        and exchanging the results with the others through the shared memory.
In the computation component, every process simulates a set of 
	runs of $\A$ using the DAGs and maintains 
	the extracted output of $\Omega$.

\subsection{DAGs}
\label{sec:dags}
\begin{figure}[tbp]
\hrule \vspace{2mm} {\small
\begin{tabbing}
 bbb\=bb\=bb\=bb\=bb\=bb\=bb\=bb \=  \kill
\> Shared variables: \\
\>\> for all $p_i\in\Pi$: $G_i$, initially empty graph\\
\\
\nnll\>  $k_i := 0$\\
\nnll\> {\bf while } $\id{true}$ {\bf do}\\
\nnll\>\> {\bf for all }  {$p_j\in\Pi$} {\bf do } $G_i \leftarrow G_i \cup G_j$\\
\nnll\>\> $d_i := $ query failure detector $\D$\\ 
\nnll\>\> $k_i := k_i+1$\\
\nnll\>\> add $[p_i,d_i,k_i]$ and 
	edges from all other vertices \\
\>\>\>\> of $G_i$ to $[p_i,d_i,k_i]$, to $G_i$
\end{tabbing}
\hrule }
\caption{\small Building a DAG: the code for each process $p_i$}
\label{fig:comm}
\end{figure}

The communication component is presented in Figure~\ref{fig:comm}.
This task maintains an ever-growing DAG that contains a finite sample of 
	the current failure detector history.
The DAG is stored in a register $G_i$ which can be updated by $p_i$ and read by all processes.

DAG $G_i$ has some special properties which follow from its construction~\cite{CHT96}.
%Consider any run of the communication component in $F$
%	with $\D$ outputting $H$.	
Let $F$ be the current failure pattern,
	and $H\in\D(F)$ be the current failure detector history. 
Then a fair run of the algorithm in Figure~\ref{fig:comm} guarantees that
	there exists a map $\tau: \Pi\times\R_{\D}\times\Nat \mapsto \Time$,
	such that, for every correct process $p_i$ and every time $t$ 
	($x(t)$ denotes here the value of variable $x$ at time $t$):
\begin{enumerate}
\item[(1)]  The vertices of $G_i(t)$ are of the form $[p_j,d,\ell]$ where $p_j\in\Pi$, 
$d\in \R_{\D}$ and ${\ell}\in \Nat$. 
\begin{enumerate}
\item[(a)] For each vertex $v=[p_j,d,{\ell}]$, $p_j \notin F(\tau(v))$ and $d=H(p_j,\tau(v))$.
That is, $d$ is the value output by $p_j$'s failure detector module at time $\tau(v)$.

\item[(b)] For each edge $(v,v')$, $\tau(v)<\tau(v')$.
That is, any edge in $G_i$ reflects the temporal order 
	in which the failure detector values are output. 
\end{enumerate}

\item[(2)] If $v=[p_j,d,{\ell}]$ and $v'=[p_j,d',{\ell}']$ are vertices of $G_i(t)$ 
	and ${\ell}<{\ell}'$ then $(v,v')$ is an edge of $G_i(t)$.
%[[PK: necessary for theorem 1
\item[(3)] $G_i(t)$ is transitively closed: if $(v,v')$ and $(v',v'')$ are edges of $G_i(t)$, 
	then $(v,v'')$ is also an edge of $G_i(t)$.
%]]
\item[(4)] For all correct processes $p_j$,  
	there is a time $t'\geq t$, a $d\in\R_{\D}$ and a 
	${\ell}\in\Nat$ such that, for every  vertex $v$ of $G_i(t)$, 
	$(v,[p_j,d,{\ell}])$ is an edge of $G_i(t')$.
\item[(5)] For all correct processes $p_j$, there is a time $t'\geq t$ such that
	$G_i(t)$ is a subgraph of $G_j(t')$.
%Let $U$ be a finite set of vertices
%and $p_i$ be any correct process in $F$. There is $d\in \R_{\D}$ and $r\in \mathbb{N}$,
%such that for every vertex $v \in \V(G)$, $(v,[p_i,d,r])\in \E(G)$.
\end{enumerate}
The properties imply that ever-growing DAGs at correct processes tend to 
	the same infinite DAG $G$: $\lim_{t\rightarrow\infty} G_i(t) = G$.
%Note that 
In a fair run of the algorithm in Figure~\ref{fig:comm}, 
	the set of processes that obtain infinitely many vertices in $G$ 
	is the set of correct processes~\cite{CHT96}.

\remove{
Note that properties (1)--(4) imply that 
for any set of vertices $V$ of $G_i(t)$,
	there is a time $t'$ such that $G_i(t')$ contains a path 
	$g$ such that every correct process appears in $g$
	arbitrarily often and $\forall v\in V$, $v\cdot g$ is also a path of $G_i(t')$.       
Furthermore, every prefix of $g$ is also a path in $G_i(t')$.
}

\subsection{Asynchronous simulation}

%As shown in~\cite{CHT96}, every DAG constructed as shown in Figure~\ref{fig:comm},
%	can be used for simulating runs of $\A$.
It is shown below that \emph{any} infinite DAG $G$ constructed as shown in Figure~\ref{fig:comm}
	can be used to simulate partial runs of $\A$
	in the \emph{asynchronous} manner:
	instead of querying $\D$, the simulation algorithm $\A'$
	uses the samples of the failure detector output captured in the DAG.  
The pseudo-code of this simulation is presented in Figure~\ref{fig:sim}. 
The algorithm is hypothetical in the sense that it uses an infinite input, 
	but this requirement is relaxed later.
%Let $\langle F,H,I,S,T \rangle$, a run of the communication component
%	described in Figure~\ref{fig:comm}, produce an infinite DAG $G$.
%Now $G$ can be used can be used to construct an \emph{asynchronous} algorithm $\A'$.

In the algorithm, each process $p_i$ is initially associated with an initial state 
	of $\A$ and performs a sequence of simulated steps of $\A$. %~\cite{Zie08}.
Every process $p_i$ maintains a shared register $V_i$ that stores the vertex of $G$
	used for the most recent step of $\A$ simulated by $p_i$.
Each time $p_i$ is about to perform a step of $\A$ it
	first reads registers $V_1,\ldots,V_n$ to obtain 
	the vertexes of $G$ used by processes $p_1,\ldots,p_n$ for simulating 
	the most recent causally preceding steps of $\A$
	(line~\ref{line:collect} in Figure~\ref{fig:sim}). 
Then $p_i$ selects the next vertex of $G$ that succeeds all vertices 
	(lines~\ref{line:repeat}-\ref{line:until}).
If no such vertex is found, $p_i$ blocks forever (line~\ref{line:wait}).  

Note that a correct process $p_i$ may block forever if $G$ contains only finitely 
	many vertices of $p_i$.
As a result an infinite run of $\A'$ may simulate an \emph{unfair} run of $\A$:
	a run in which some correct process takes only finitely many steps. 
But every finite run simulated by $\A'$ is a partial run of $\A$. 

\begin{figure}[tbp]
\hrule \vspace{2mm} {\small
\begin{tabbing}
 bbb\=bb\=bb\=bb\=bb\=bb\=bb\=bb \=  \kill
\> Shared variables: \\
\>\> $V_1,\ldots, V_n := \bot,\ldots,\bot$, \\
\>\>\>\> \{for each $p_j$, $V_j$ is the vertex of $G$\\ 
\>\>\>\> corresponding to the latest simulated step of $p_j$\}\\
\>\> Shared variables of $\A$\\
\\
%\> initialize the state of $\A$ with the input value\\
\nnll\> initialize the simulated state of $p_i$ in $\A$, based on $I'$ \\
%\>\>\> simulation state of $p_i$ in $I'$\\  
\nnll\> $\ell :=0$\\
\nnll\> \textbf{while} $\true$ \textbf{do}\\
\>\> \{Simulating the next $p_i$'s step of $\A$\}\\
\nnll\label{line:collect}\>\> $U:= [V_1,\ldots,V_n]$\\
\nnll\label{line:repeat}\>\> {\bf repeat}\\
\nnll\>\>\>  $\ell := \ell+1$\\	 
\nnll\label{line:wait}\>\>\> {\bf wait until} $G$ includes $[p_i,d,\ell]$ for some $d$\\
\nnll\label{line:until}\>\> {\bf until} $\forall j$, %\in\{1,\ldots,i-1,i+1,\ldots,n\}$,  
	$U[j]\neq \bot$: $(U[j],[p_i,d,\ell]) \in G$\\
% \hspace{1cm} \{ Wait until all observed steps of other processes are in the past\}\\
\nnll\label{line:upd}\>\> $V_i:= [p_i,d,\ell]$\\
\nnll\label{line:sim}\>\> take the next $p_i$'s step of $\A$ using $d$ as the output of $\D$
\end{tabbing}
\hrule }
\caption{\small DAG-based asynchronous algorithm $\A'$: code for each $p_i$}
\label{fig:sim}
\end{figure}

\begin{theorem}
\label{th:sim1}
Let $G$ be the DAG produced in a fair run $R=\langle F,H,I,S,T \rangle$ 
	of the communication component in Figure~\ref{fig:comm}.
Let $R'=\langle F',H',I',S',T' \rangle$ be any fair run of $\A'$ using $G$.
Then the sequence of steps simulated by $\A'$ in $R'$ 
	belongs to a (possibly unfair) run of $\A$, $R_{\A}$, with input vector of $I'$ and 
	failure pattern $F$. 
Moreover, the set of processes that take infinitely many steps in $\R_{\A}$ 
	is $\correct(F)\cap\correct(F')$, and if $\correct(F)\subseteq \correct(F')$, 
	then $R_{\A}$ is fair. 
\end{theorem}
\begin{proof}
Recall that a step of a process $p_i$ can be either a \emph{memory} step in which $p_i$ 
	accesses shared memory or a \emph{query} step in which $p_i$ 
	queries the failure detector.
Since memory steps simulated in $\A'$ are performed as in $\A$, 
	to show that algorithm $\A'$ indeed simulates a run of $\A$ with failure pattern 
	$F$, it is enough to make sure that the sequence of 
	simulated \emph{query} steps in the simulated run 
	(using vertices of $G$) \emph{could have been observed} 
	in a run $R_{\A}$ of $\A$ with failure pattern $F$ and the input vector based on $I'$.

Let $\tau$ be a map associated with $G$ that carries each vertex of $G$ to
	 an element in $\Time$ such that (a) for any vertex $v=[p,d,\ell]$ of $G$, 
	$p \notin F(\tau(v))$ and $d=H(p_j,\tau(v))$, and (b) for every edge $(v,v')$ of $G$, 
	$\tau(v)<\tau(v')$ (the existence of $\tau$ 
	is established by property (5) of DAGs in Section~\ref{sec:dags}).
For each step $s$ simulated by $\A'$ in $\R'$, let $\tau'(s)$ denote time when step $s$ \emph{occurred} 
	in $\R'$, i.e., 
	when the corresponding line~\ref{line:sim} in Figure~\ref{fig:sim} was executed,
	and $v(s)$ be the vertex of $G$ used for simulating $s$, i.e., the value of $V_i$ when 
	$p_i$ simulates $s$ in line~\ref{line:sim} of Figure~\ref{fig:sim}.

Consider query steps $s_i$ and $s_j$ simulated by processes $p_i$ and $p_j$, respectively.  
Let $v(s_i)=[p_i,d_i,\ell]$ and $v(s_j)=[p_j,d_j,m]$.
WLOG, suppose that $\tau([p_i,d_i,\ell])<\tau([p_j,d_j,m])$, i.e., $\D$ outputs $d_i$ at $p_i$ 
	before outputting $d_j$ at $p_j$.

If $\tau'(s_i)<\tau'(s_j)$, i.e., $s_i$ is simulated by $p_i$ before $s_j$ is simulated by $p_j$,  
	then the order in which $s_i$ and $s_j$ see value $d_i$ and $d_j$ is the run produced by $\A'$ 
	is consistent with the output of $\D$, i.e., the values $d_i$ and $d_j$ indeed
	could have been observed in that order.

Suppose now that $\tau'(s_i)>\tau'(s_j)$.
If $s_i$ and $s_j$ are not causally related in the simulated run, 
	then $R'$ is indistinguishable from 
	a run in which $s_i$ is simulated by $p_i$ \emph{before} $s_j$ is simulated by $p_j$.
Thus, $s_i$ and $s_j$ can still be observed in a run of $A$.

Now suppose, by contradiction that $\tau'(s_i)>\tau'(s_j)$ and 
	$s_j$ causally precedes $s_i$ in the simulated run, 
	i.e., $p_j$ simulated at least one write step $s_j'$ after $s_j$,
	and $p_i$ simulated at least one read step $s_i'$ before $s_i$, such that 
	$s_j'$ took place before $s_i'$ in $R'$. 
Since before performing the memory access of $s_j'$, $p_j$ 
	updated $V_j$ with a vertex $v(s_j')$ that succeeds $v(s_j)$ in $G$ (line~\ref{line:upd}), 
	and $s_i'$ occurs in $R'$ after $s_j'$,
	$p_i$ must have found $v(s_j')$ or a later 
	vertex of $p_j$ in $V_j$ before simulating step $s_i$
	(line~\ref{line:collect}) and, thus, 
	the vertex of $G$ used for simulating $s_i$ must be a descendant of $[p_j,d_j,m]$, and,
	by properties (1) and (3) of DAGs (Section~\ref{sec:dags}), 
	$\tau([p_i,d_i,\ell])>\tau([p_j,d_j,m])$
	--- a contradiction.
Hence, the sequence of steps of $\A$ simulated in $R'$ could have been observed 
	in a run $R_{\A}$ of $\A$ with failure pattern $F$.

Since in $\A'$, a process simulates only its own steps of $\A$,
	every process that appears infinitely often in $R_{\A}$ is 
	in  $\correct(F')$.
Also, since each faulty in $F$ process contains only finitely many vertices in $G$,
	eventually, each process in $\correct(F')-\correct(F)$ is blocked 
	in line~\ref{line:wait} in Figure~\ref{fig:sim},
	and, thus, every process that appears infinitely often in $R_{\A}$ is also
	in  $\correct(F)$.
Now consider a process $p_i\in\correct(F')\cap\correct(F)$.
Property (4) of DAGs implies that for every set $V$ of vertices of $G$, 
	there exists a vertex of $p_i$ in $G$
	such that for all $v'\in V$, $(v',v)$ is an edge in $G$.
Thus, the wait statement in line~\ref{line:wait} cannot block $p_i$ forever,
	and $p_i$ takes infinitely many steps in $R_{\A}$.

Hence, the set of processes that appear infinitely often in $R_{\A}$ is exactly $\correct(F')\cap\correct(F)$.
Specifically, if $\correct(F)\subseteq \correct(F')$, then the set of 
	processes that appear infinitely often in $R_{\A}$ is $\correct(F)$, 
	and the run is fair.
\end{proof}
Note that in a fair run, the properties of the algorithm in Figure~\ref{fig:sim} remain 
	the same if the infinite DAG $G$ is replaced 
	with a finite ever-growing DAG $\bar G$ constructed in parallel (Figure~\ref{fig:comm}) 
	such that $\lim_{t\to\infty} \bar G = G$. 
This is because such a replacement only affects the wait statement in line~\ref{line:wait} which  blocks $p_i$
	until the first vertex of $p_i$ that causally succeeds every simulated step recently "witnessed" by $p_i$
	is found in $G$, 
%	to have been taken since the last simulated step of $p_i$ appears in $G_i$,
	%such a vertex is eventually present in $\bar G$ if $p_i$ is correct 
	but this cannot take forever if $p_i$ is correct
	(properties (4) and (5)	of DAGs in Section~\ref{sec:dags}).
The wait blocks forever if the vertex is absent in $G$, which may happen only if $p_i$ is faulty. 

\subsection{BG-simulation}
\label{sec:bg}

Borowsky and Gafni proposed in~\cite{BG93b,BGLR01}, a simulation technique
by which $k+1$ \emph{simulators} $q_1,\ldots,q_{k+1}$ can wait-free simulate a $k$-resilient 
execution of any asynchronous $n$-process protocol.
Informally, the simulation works as follows.
Every process $q_i$ tries to simulate steps of all $n$ processes $p_1,\ldots,p_n$
	in a round-robin fashion.   
Simulators run an \emph{agreement protocol} to make sure that 
	every step is simulated at most once. 
Simulating a step of a given process may block forever if and only if 
	some simulator has crashed in the middle of the corresponding agreement protocol.
Thus, even if $k$ out of $k+1$ simulators crash, at least $n-k$ simulated processes 
	can still make progress.
The simulation thus guarantees at least $n-k$ 
	processes in $\{p_1,\ldots,p_n\}$ accept infinitely many simulated steps.

\begin{figure}[htbp]
  \centering
  \includegraphics[scale=0.6]{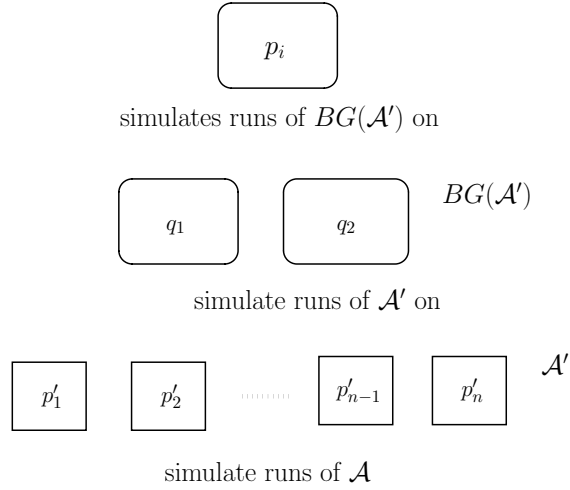}
  \caption{\small Three levels of simulation: real processes $p_i$ simulate a system of two 
	BG-simulators $q_1$ and $q_2$
	that run $BG(\A')$ to simulate an $(n-1)$-resilient run of $\A'$ on $p_1',\ldots,p_n'$.}
  \label{fig:triple}
\end{figure}

In the computational component of the reduction algorithm, the BG-simulation technique is used as follows. 
Let $BG(\A')$ denote the simulation protocol for $2$ processes $q_1$ and $q_{2}$
	which allows them to simulate, in a wait-free manner, a $1$-resilient 
	execution of algorithm $\A'$ for $n$ processes $p_1,\ldots,p_n$. 
The complete reduction algorithm  thus employs a \emph{triple} simulation (Figure~\ref{fig:triple}): 
	every process $p_i$ simulates multiple runs of two processes $q_1$ and $q_{2}$ that
	use BG-simulation to produce a $1$-resilient run of $\A'$ on processes $p_1',\ldots,p_n'$ 
	in which steps of the original algorithm $\A$ are periodically simulated
	using (ever-growing) DAGs $G_1,...,G_n$.
(To avoid confusion, we use $p_j'$ to denote the process that models $p_j$ in a 
	run of $\A'$ simulated by a ``real'' process $p_i$.)

We are going to use the following property which is trivially satified by BG-simulation:

\begin{enumerate}
\item[(BG0)] A run of BG-simulation in which every simulator take infinitely many steps 
simulates a run in which every simulated process takes infinitely many steps.
\end{enumerate}

\subsection{Using consensus}
\label{sec:cons}

The triple simulation we are going to employ faces one complication though. 
The simulated runs of the asynchronous algorithm $\A'$ may vary depending    
	on which process the simulation is running.
This is because $G_1,...,G_n$ are maintained by a parallel computation component (Figure~\ref{fig:comm}),
	and a process simulating a step of $\A'$ may perform a different number of cycles reading 
	the current version of its DAG before
	a vertex with desired properties is located (line~\ref{line:wait} in Figure~\ref{fig:sim}). 
	%, depending on time when the step is simulated.
Thus, the same sequence of steps of $q_1$ and $q_2$ simulated at different processes
	may result in different $1$-resilient runs of $\A'$: waiting until a vertex 
	$[p_i,d,\ell]$ appears in $G_j$ at process $p_j$ may take different number 
	of local steps checking $G_j$,
	depending on the time when $p_j$ executes the wait statement in line~\ref{line:wait} 
	of Figure~\ref{fig:sim}.
\begin{figure}[tbp]
\hrule \vspace{2mm} {\small
\begin{tabbing}
 bbb\=bb\=bb\=bb\=bb\=bb\=bb\=bb \=  \kill
\>$r := 0$\\ 
\>{\bf repeat}\\
\>\>  $r := r+1$\\
\>\>  \textbf{if} $G$ contains $[p_i,d,\ell]$ for some $d$ \textbf{then} $u := 1$\\
\>\> \textbf{else} $u := 0$\\	 
\>\> $v := \textsf{cons}^{i,\ell}_r.\textit{propose}(u)$\\
\>\textbf{until} $v=1$
\end{tabbing}
\hrule }
\caption{\small Expanded line~\ref{line:wait} of Figure~\ref{fig:sim}: waiting until $G$ includes 
	a vertex $[p_i,d,\ell]$ 
for some $d$. Here $G$ is any DAG generated by the algorithm in Figure~\ref{fig:comm}.}
\label{fig:wait}
\end{figure}

To resolve this issue,  the wait statement is implemented 
	using a series of consensus instances $\textsf{cons}^{i,\ell}_1$, $\textsf{cons}^{i,\ell}_2$, 
	$\ldots$ (Figure~\ref{fig:wait}).
If $p_i$ is correct, then eventually each correct process will have  
	a vertex $[p_{i},d,\ell]$ in its DAG and, thus, the code in Figure~\ref{fig:wait} 
	is non-blocking, and Theorem~\ref{th:sim1} still holds.
Furthermore, the use of consensus ensures that if a process, while simulating a step of $\A'$ 
	at process $p_i$, 
	went through $r$ steps before reaching line~\ref{line:until} in Figure~\ref{fig:sim},
	then every process simulating this very step does the same.
Thus, a given sequence of steps of $q_1$ and $q_2$ 
	will result in the same simulated $1$-resilient run of $\A'$,
	regardless of when and where the simulation is taking place.

\subsection{Extracting $\Omega$}
\label{sec:omega}

The computational component of the reduction algorithm is presented in Figure~\ref{fig:red}.
In the component, every process $p_i$ locally simulates multiple runs of a system of
	$2$ processes $q_1$ and $q_{2}$ that run algorithm $BG({\A'})$,
	to produce a $1$-resilient run of $\A'$ 
	(Figures~\ref{fig:sim} and~\ref{fig:wait}). 
Recall that $\A'$, in its turn, simulates a run of the original algorithm $\A$, 
	using, instead of $\D$, the values provided by 
	an ever-growing DAG $G$.
In simulating the part of $\A'$ of process $p_i'$ presented in Figure~\ref{fig:wait}, 
	$q_1$ and $q_2$ count each access of a consensus instance $\textsf{cons}^{i,\ell}_r$ 
	as \emph{one local step} of $p_i'$ that need to be simulated.
Also, in $BG(\A')$, when $q_j$ is about to simulate the first step of $p_i'$, 
	$q_j$ uses its own input value as an input value of $p_i'$.

For each simulated state $S$ of $BG({\A'})$, 
	$p_i$ periodically checks whether the state of $\A$ in $S$ 
	is \emph{deciding}, i.e., whether some process has decided 
	in the state of $\A$ in $S$.
As we show, eventually, the same infinite non-deciding $1$-resilient run of $\A'$ will be simulated 
	by all processes, which allows for extracting the output of $\Omega$.

The algorithm in Figure~\ref{fig:red} explores \emph{solo} extensions of $q_1$ and $q_2$ 
	starting from growing prefixes.  
Since, by property (BG0) of BG-simulation (Section~\ref{sec:bg}), 
	a run of $BG(\A')$ in which both $q_1$ and $q_2$ participate infinitely often 
	simulates a run of $\A'$ in which every $p_j\in\{p_1',\ldots,p_n'$ 
	participates infinitely often,	and, by Theorem~\ref{th:sim1}, such a run
	will produce a fair and thus deciding run of $\A$.
Thus, if there is an infinite non-deciding run simulated by the algorithm in  Figure~\ref{fig:sim},
	it must be a run produced by a solo extension of $q_1$ or $q_2$ starting from some
	finite prefix.  

\begin{figure}[tbp]
\hrule \vspace{2mm} {\small
\begin{tabbing}
 bbb\=bb\=bb\=bb\=bb\=bb\=bb\=bb \=  \kill
%\> \\
\nnll\> \textbf{for all} binary $2$-vectors $J_0$ \textbf{do} \\
\>\>\> \{ For all possible consensus inputs for $q_1$ and $q_2$ \} \\
%\nnll\>\> $S_0 := $ the initial state of $BG_{\A'}$ with input vector $I$\\ 
\nnll\>\> $\sigma_0 := $ the empty string\\
\nnll\label{line:explore0}\>\> \textit{explore}$(J_0,\sigma_0)$\\
\\
\nnll\>\textbf{function} \textit{explore}$(J,\sigma)$ \\
\nnll\>\> \textbf{for all} $q_j=q_1,q_{2}$ \textbf{do}\\
\nnll\label{line:solo0}\>\>\> $\rho:=$ empty string\\
\nnll\label{line:solo1}\>\>\> \textbf{repeat} \\
\nnll\>\>\>\> $\rho:= \rho\cdot q_j$\\
\nnll\>\>\>\> let $p_{\ell}'$ be the process that appears the least 
		in $\textit{SCH}_{\A'}(J,\sigma\cdot\rho)$\\ 
\nnll\label{line:output}\>\>\>\> $\Omega\id{-output} := p_{\ell}$\\
\nnll\label{line:solo2}\>\>\> \textbf{until} $\textit{ST}_{\A}(J,\sigma\cdot\rho)$ is decided\\
%\nnll\>\> \textbf{for all} $q_j=q_1,q_{2}$ \textbf{do}\\
%\nnll\>\> $S' := $ the state of $BG_{\A'}$ resulting after applying \\
%\>\>\>\> a step of $q_j$ to $S$\\ 
\nnll\label{line:explore1}\>\> \textit{explore}$(J,\sigma\cdot q_1)$\\
\nnll\label{line:explore2}\>\> \textit{explore}$(J,\sigma\cdot q_2)$
\end{tabbing}
\hrule }
\caption{\small Computational component of the reduction algorithm: code for each process $p_i$. 
Here $\textit{ST}_{\A}(J,\sigma)$ denotes the state of $\A$ reached by the partial run of $\A'$ 
	simulated in the partial run of $BG({\A'})$ with schedule 
	$\sigma$ and input state $J$, and $\textit{SCH}_{\A'}(J,\sigma)$ denotes 
	the corresponding schedule of $\A'$.}
\label{fig:red}
\end{figure}

\begin{lemma}
\label{lem:nondec}
The algorithm in Figure~\ref{fig:red} eventually forever executes lines~\ref{line:solo1}--\ref{line:solo2}.
\end{lemma}
\begin{proof}
Consider any run of the algorithm in Figures~\ref{fig:comm}, \ref{fig:wait} and~\ref{fig:red}.
Let $F$ be the failure pattern of that run.
Let $G$ be the infinite limit DAG approximated by the algorithm in Figure~\ref{fig:comm}. 
By contradiction, suppose that lines~\ref{line:solo1}--\ref{line:solo2} in Figure~\ref{fig:red} 
	never block $p_i$.

Suppose that for some initial $J_0$, the call of \textit{explore}$(J_0,\sigma_0)$ performed by $p_i$ 
	in line~\ref{line:explore0} never returns.
Since the cycle in lines ~\ref{line:solo1}--\ref{line:solo2} in Figure~\ref{fig:red} 
	always terminates, 
	there is an infinite sequence of recursive calls \textit{explore}$(J_0,\sigma_0)$,
\textit{explore}$(J_0,\sigma_1)$, \textit{explore}$(J_0,\sigma_2)$, $\ldots$, where 
each $\sigma_{\ell}$ is a one-step extension of $\sigma_{\ell-1}$. 
Thus, there exists an infinite never deciding schedule $\tilde \sigma$ such that the run of $BG(\A')$ 
	based on $\tilde \sigma$ and $J_0$ produces a never-deciding run of $\A'$.
Suppose that both $q_1$ and $q_2$ appear in $\tilde\sigma$ infinitely often.
By property (BG0) of BG-simulation (Section~\ref{sec:bg}), 
	a run of $BG(\A')$ in which both $q_1$ and $q_2$ participate infinitely often 
	simulates a run of $\A'$ in which every $p_j\in\{p_1',\ldots,p_n'\}$ 
	participates infinitely often,	and, by Theorem~\ref{th:sim1}, such a run
	will produce a fair and thus deciding run of $\A$ --- a contradiction.

Thus, if there is an infinite non-deciding run simulated by the algorithm in  Figure~\ref{fig:sim},
	it must be a run produced by a solo extension of $q_1$ or $q_2$ starting from some
	finite prefix.
Let $\bar \sigma$ be the first such prefix in the order defined by the algorithm in Figure~\ref{fig:sim}
	and $q_{\ell}$ be the first process whose solo extension of $\sigma$ is never deciding.
Since the cycle in lines ~\ref{line:solo1}--\ref{line:solo2} 
	always terminates, the recursive exploration of finite prefixes $\sigma$ in 
	lines~\ref{line:explore1} and~\ref{line:explore2} eventually reaches $\bar\sigma$, 
	the algorithm reaches line~\ref{line:solo0}
	with $\sigma=\bar\sigma$ and $q_j=q_{\ell}$. 
Then the succeeding cycle in lines ~\ref{line:solo1}--\ref{line:solo2} never terminates --- a contradiction. 

Thus, for all inputs $J_0$,  the call of \textit{explore}$(J_0,\sigma_0)$ performed by $p_i$ 
	in line~\ref{line:explore0} returns.
Hence, for every finite prefix $\sigma$, any solo extension of $\sigma$ produces a finite 
	   deciding run of $\A$.
We establish a contradiction, by deriving a wait-free  algorithm that solves consensus 
	among $q_1$ and $q_2$. 

Let $\tilde G$ be the infinite limit DAG constructed in Figure~\ref{fig:comm}. 
Let $\beta$ be a map from vertices of $\tilde G$ to $\Nat$ defined as follows:
	for each vertex $[p_i,d,\ell]$ in $G$, $\beta([p_i,d,\ell])$ 
	is the value of variable $r$ at the moment when any run of $\A'$ (produced 
	by the algorithm in~Figure~\ref{fig:sim}) 
	exits the cycle in Figure~\ref{fig:wait}, while waiting until $[p_i,d,\ell]$ appears in $G$.
If there is no such run, $\beta([p_i,d,\ell])$ is set to $0$. 
Note that the use of consensus implies that if in any simulated run of $\A'$, 
	$[p_i,d,\ell]$ has been found after $r$ 
	iterations, then $\beta([p_i,d,\ell])=r$, i.e., $\beta$ is well-defined. 

Now we consider an asynchronous read-write algorithm $\A'_{\beta}$ that is defined exactly like $\A'$,
	but instead of going through the consensus invocations in Figure~\ref{fig:wait},
	$\A'_{\beta}$ performs $\beta([p_i,d,\ell])$ \emph{local} steps.
Now consider the algorithm $BG(\A'_{\beta})$ that is defined exactly as $BG(\A')$ except that
	in $BG(\A'_{\beta})$, $q_1$ and $q_2$ BG-simulate runs of $\A'_{\beta}$.
For every sequence $\sigma$ of steps of $q_1$ and $q_2$, the runs of $BG(\A')$ 
	and $BG(\A'_{\beta})$ agree on the sequence of steps of $p_1',\dots,p_n'$ 
	in the corresponding runs of $\A'$ and $\A'_{\beta}$, respectively.
Moreover, they agree on the runs of $\A$ resulting from these runs of $\A'$ and $\A'_{\beta}$.
This is because the difference between $\A'$ and $\A'_{\beta}$ consist only in 
	the local steps and does not affect the simulated state of $\A$.

We say that a sequence $\sigma$ of steps of $q_1$ and $q_2$ is \emph{deciding with $J_0$}, if, 
	when started with $J_0$, the run of $BG(\A'_{\beta})$ produces a deciding run of $\A$.  
By our hypothesis, every eventually solo schedule $\sigma$ is deciding for each input $J_0$.
As we showed above, every schedule in which both $q_1$ and $q_2$ 
	appear sufficiently often 
	is deciding by property (BG0) of BG-simulation.
Thus, every schedule of $BG(\A'_{\beta})$ is deciding for all inputs.

Consider the trees of all deciding schedules of $BG(\A'_{\beta})$ for all possible inputs $J_0$.
All these trees have finite branching (each vertex has at most $2$ descendants) and finite paths. 
By K\"onig's lemma, the trees are finite.
Thus, the set of vertices of $\tilde G$ used by the runs of $\A'$ simulated by deciding 
	schedules of $BG(\A'_{\beta})$ is also finite.
Let $\bar G$ be a finite subgraph of $\tilde G$ that includes all vertices of $\tilde G$ used
	by these runs.

Finally, we obtain a wait-free consensus algorithm  for $q_1$ and $q_2$ that works as follows. 
Each $q_j$ runs $BG(\A'_{\beta})$ (using a finite graph $\bar G$) 
	until a decision is obtained in the simulated run of $\A$. 
At this point, $q_j$ returns the decided value.
But $BG(\A'_{\beta})$ produces only deciding runs of $\A$, and  
	each deciding run of $\A$ solves consensus for inputs provided by $q_1$ and $q_2$ ---
	a contradiction. 
\end{proof}

\begin{theorem}
\label{th:nec}
In all environments $\E$, if a failure detector $\D$ can be used to solve consensus in $\E$, then 
	$\Omega$ is weaker than $\D$ in $\E$.
\end{theorem}
\begin{proof}
Consider any run of the algorithm in Figures~\ref{fig:comm}, \ref{fig:wait} and~\ref{fig:red}
	with failure pattern $F$.

By Lemma~\ref{lem:nondec}, at some point, every correct process $p_i$ gets stuck
	in lines~\ref{line:solo1}--\ref{line:solo2} simulating longer and longer $q_j$-solo extension of 
	some finite schedule $\sigma$ with input $J_0$.
Since, processes $p_1,\ldots,p_n$  use a series of consensus instances to simulate runs 
	of $\A'$ in exactly the same way,
	the correct processes eventually agree on $\sigma$ and $q_j$. 

Let $e$ be the sequence of process identifiers in the $1$-resilient execution
	of $\A'$ simulated by $q_1$ and $q_2$ in schedule $\sigma\cdot(q_j)$ with input $J_0$.
Since a $2$-process BG-simulation produces a $1$-resilient run of $\A'$, 
	at least $n-1$ simulated processes in $p_1',\ldots,p_n'$ appear in $e$
	infinitely often.
Let $U$ ($|U|\geq n-1$) be the set of such processes. 

Now we show that exactly one correct (in $F$) process appears in $e$ only finitely often.
Suppose not, i.e., $\correct(F)\subseteq U$.
By Theorem~\ref{th:sim1}, the run of $\A'$ simulated a far run of $\A$, 
	and, thus, the run must be deciding --- a contradiction.
Since $|U|\geq n-1$, exactly one process appears in the run of $\A'$ only finitely often. 
Moreover, the process is correct.

Thus, eventually, the correct processes in $F$ stabilize at simulating longer and longer prefixes 
	of the same infinite non-deciding $1$-resilient run of $\A'$.
Eventually, the same correct process will be observed to take the least number of steps in the run
	and output in line~\ref{line:output}
	 --- the output of $\Omega$ is extracted.	
\end{proof}

%=====================================================================
\section{Related Work} 
\label{sec:related}
%=====================================================================

Chandra et al. derived the first ``weakest failure detector'' result
	by showing that $\Omega$ is necessary to solve consensus in the message-passing model
	in their fundamental paper~\cite{CHT96}.
%At every process and each time, $\Omega$ outputs a \emph{leader} process identifier 
%	and guarantees that, eventually, all processes agree on the same correct leader. 
The result was later generalized to the read-write shared memory model~\cite{LH94,GK08}.
\footnote{The result for the shared memory as stated in~\cite{LH94}, 
	but the only published proof of it appears in~\cite{GK08}.}

The technique presented in this paper builds atop two fundamental results.
The first is the celebrated BG-simulation~\cite{BG93b,BGLR01} that 
	allows $k+1$ processes simulate, in a wait-free manner, a $k$-resilient run of any $n$-process 	asynchronous algorithm.
The second is a brilliant observation made by Zieli\'nski~\cite{Zie08} that 
	any run of an algorithm $\A$ using a failure detector $\D$ induces an \emph{asynchronous} 
	algorithm that simulates (possibly unfair) runs of $\A$.
The recursive structure of the algorithm in Figure~\ref{fig:red} is also borrowed from~\cite{Zie08}. 
Unlike~\cite{Zie08}, however, the reduction algorithm  of this paper assumes the conventional read-write memory model
	without using immediate snapshots~\cite{BG93a}.
Also, instead of growing "precedence" and "detector" maps of~\cite{Zie08},
	this paper uses directed acyclic graphs \'a la~\cite{CHT96}.
A (slightly outdated) survey on the literature on  failure detector is presented in~\cite{FGK11}.  

%=====================================================================
\section{Concluding Remarks}
\label{sec:discussion}
%=====================================================================

This paper presents another proof that $\Omega$ is 
	the weakest failure detector 
	to solve consensus in read-write shared memory models.  
The proof applies a novel reduction technique, and is based on the very fact 
	that wait-free $2$-process consensus is impossible,
	unlike the original technique of~\cite{CHT96}
	that partially rehashes elements of the consensus impossibility
	proof.

%Our technoque can be used to show that $\Omega$ is necessary to solve any 
%\emph{$1$-resilient impossible} distributed colorless task~\cite{HS99}, %
%	i.e., a task that cannot be solved if at least one process 
%	may fail.	
%Indeed, every failure detector that circumvents such an impossiblity,
%	can be used to solve consensus~\cite{RAchid}.   

A related problem is determining the weakest failure detector 
	for a generalization of consensus,  
	$(n,k)$-set agreement, in which  
	$n$ processes have to decide on at most $k$ distinct proposed 
	values. %~\cite{RT06,CZCL07,GHKLN07}.
The weakest failure detector for $(n,1)$-set agreement  (consensus)
	is $\Omega$.
For $(n,n-1)$-set agreement (sometimes called simply 
	set agreement in the literature), it
	is anti-$\Omega$, a failure detector that outputs, when queried, a process
	identifier, so that some correct process identifier is output
	only finitely many times~\cite{Zie10}.

Finally, the general case of $(n,k)$-set agreement was resolved using 
an elaborated and extended version of the technique proposed in this
paper~\cite{GK11}. 
Intuitively, BG simulation allows $k+1$ processes to simulate a $k$-resilient 
	run of any asynchronous algorithm, and, generalizing 
	the technique described 
	in this paper, we can derive an infinite
	non-deciding $k$-resilient  
	run $R$ of $\A'$. 
At least one correct process appears only finitely often in $R$ (otherwise,
	the run would be deciding).
Thus, a failure detector that periodically outputs the latest $n-k$ processes 
	in growing prefixes of $R$ 
	guarantees that eventually some correct process is never output.    
It can be easily shown that this information about failures 
	is sufficient to solve $(n,k)$-set agreement.  
For $k>1$, we cannot use consensus to make sure that correct processes simulate 
	runs of $\A'$ in exactly the same way, regardless of how their local DAGs evolve.
Therefore, our generalized reduction algorithm employs a slightly more sophisticated 
	``eventual agreement'' mechanism to make sure that the simulation converges.

%\section*{Acknowledgements}

%I am grateful to Eli Gafni for the crucial observation that
%	a reduction algorithm can directly use consensus objects
%	to help processes agree on the simulated executions of $\A'$. 
%to Rachid Guerraoui for indirect support. 
% the "unaware muse"~\cite{AGRRT06} of this work. 
%The hierarchy of tasks described in Section~\ref{sec:discussion}
%	is a "folklore" observation made by many researchers, including Eli Gafni, Rachid Guerraoui, 
%	Achour Mostefaoui, and Michel Raynal.

{\small 
\bibliography{references}
}

\end{document}